\documentclass[a4paper,aps,onecolumn,showpacs,showkeys,nofootinbib,preprintnumbers,superscriptaddress,amsmath,amssymb,amsfonts]{revtex4}
\usepackage{graphicx}
\usepackage{bm,natbib}
\usepackage{amsmath}
\usepackage[english]{babel}
\usepackage[T1]{fontenc}
\usepackage{amssymb}
\usepackage{amsfonts}

\usepackage{amsthm}                      

\usepackage{braket}
\usepackage{quoting}
\usepackage{comment}
\quotingsetup{font=small}
\usepackage{csquotes}
\usepackage{url}

\usepackage{epsfig}
\usepackage{colordvi}
\usepackage{psfrag}
\usepackage{color}
\usepackage{mathrsfs}
\usepackage{dcolumn}
\usepackage{multirow}
\usepackage{hyperref}
\usepackage{epstopdf}
\usepackage{bm}
\usepackage{times}

\hypersetup{
                    colorlinks=true,        
                    linkcolor=blue,         
                    citecolor=magenta,      
}

\begin{document}
\title{ Non-Local Curvature Gravity Cosmology via Noether Symmetries}


\author{Adriano Acunzo}
\email{adriano.acunzo@unina.it}
\affiliation{Department of Physics ``E. Pancini'', University of Naples ``Federico II'', Naples, Italy.}

\author{Francesco Bajardi}
\email{francesco.bajardi@unina.it}
\affiliation{Department of Physics ``E. Pancini'', University of Naples ``Federico II'', Naples, Italy.}
\affiliation{INFN Sez. di Napoli, Compl. Univ. di Monte S. Angelo, Edificio G, Via Cinthia, I-80126, Naples, Italy.}

\author{Salvatore Capozziello}
\email{capozziello@unina.it}
\affiliation{Department of Physics ``E. Pancini'', University of Naples ``Federico II'', Naples, Italy.}
\affiliation{INFN Sez. di Napoli, Compl. Univ. di Monte S. Angelo, Edificio G, Via Cinthia, I-80126, Naples, Italy.}
\affiliation{Scuola Superiore Meridionale, Largo San Marcellino 10, I-80138, Naples, Italy.}
\affiliation{Laboratory for Theoretical Cosmology, Tomsk State University of Control Systems and Radioelectronics (TUSUR), 634050 Tomsk, Russia.}
\affiliation{Department of Mathematics, Faculty of Civil Engineering,
VSB-Technical University of Ostrava, Ludvika Podeste 1875/17, 
708 00 Ostrava-Poruba, Czech Republic.}

\noaffiliation  

\date{\today}

\begin{abstract}
We consider  extensions of General Relativity based on the non-local function $f(R, \Box^{-1} R)$, where $R$ is the Ricci curvature scalar and the non-locality is due to the term $\Box^{-1} R$. We focus on  cosmological minisuperspaces and select viable models by the Noether Symmetry Approach. Then we find viable  exact solutions pointing out the role of non-locality in cosmology.
\end{abstract}


\maketitle


\section{Introduction}

 Einstein's General Relativity (GR) gives the best description of gravitational interaction. It was proposed in 1915 and gained a wide success thanks to several experiments and observations which continuously  confirm its validity. These confirmations occurred at several scales of energy, ranging from  Solar System up to  cosmology. For instance, the expansion of the universe and the description of large scale structure were a formidable probes at cosmological level. Furthermore, the detection of gravitational waves and the final evidences of black holes existence provided exceptional test-beds for the theory.

Nevertheless,  several shortcomings arose both at ultraviolet (UV) and infrared (IR) scales questioning the validity of GR as the final theory of gravity. In other words,  they  suggest that GR could be an effective theory not working at any energy scale. 

A first type of shortcomings is represented by the space-time singularities. For example,  the Kruskal-Szekeres maximal extension of the Schwarzschild solution exhibits a true space-time singularity at the gravitational center, which cannot be removed by  coordinate transformations. At the astrophysical level, the so called ``Galaxy rotation curve problem'' occurs \cite{Babocock, Bosma:1981zz}. It consists of the discrepancy between the theoretical and the observed speed of stars orbiting around galaxies. In order to properly address this issue, Dark Matter was introduced. It represents a hypothetical fluid supposed to account for the 26\% of the Universe content.

Similar problems are suffered by standard cosmology. Current observations show an accelerated expansion of the Universe at  late times. This effect can be taken into account by introducing in the Einstein-Hilbert action the cosmological constant $\Lambda$. It can be physically interpreted as an anomalous  fluid with negative pressure, dubbed Dark Energy, which should drive the cosmological expansion~\cite{Peebles:2002gy, Padmanabhan:2002ji}. However, so far, there is no  final experimental evidence explaining  Dark Matter and Dark  Energy at fundamental level as further particles beyond the Standard Model. The puzzling situation is that the  $\Lambda$CDM model is compatible with precision cosmology observations \cite{Spergel:2003cb} but it suffers several conceptual shortcomings  at theoretical level. For instance, the observed value of $\Lambda$ is 120 orders of magnitude lower than the vacuum energy density predicted by Quantum Field Theory (QFT) in curved space-times.

Regarding UV scales, GR turns out to be renormalizable up to  second-loop level \cite{Goroff:1985th}, which means that incurable divergences arise in the attempt to merge GR with Quantum Mechanics. Furthermore, GR cannot be treated under the standard of  Yang-Mills theories, so that it cannot be unified to the other fundamental interactions. Furthermore, the absence of a Hilbert space and the lack of a probabilistic interpretation of the wave function yield several obstacles so far unsolved. In other words, a quantum theory of gravity is not yet available at the moment.  

These problems suggest that GR needs to be generalized to a more complete theory capable of overcoming low and high-energy issues. In this framework, extended theories of gravity (ETGs) can be taken into account \cite{Capozziello:2011et}.  They introduce higher order terms in the curvature invariants or non-minimal couplings between scalar field and geometry into the Einstein-Hilbert action. These additional terms come from the one-loop effective action predicted by the semi-classical approach of QFT on curved space-time \cite{Birrell:book, Adams:1990pn, Cotsakis:2006zn, Amendola:1993bg}. In this approach, metric is considered as a classical field, while  matter is treated as quantum fields. It comes out that the effective  Lagrangian contains geometrical UV-divergent terms, proportional to $R^2$,   $R^{\mu\nu}R_{\mu\nu}$ and other curvature invariants \cite{Birrell:book}.
 See also \cite{Capozziello:2011et, Capozziello:2019klx,  DeFelice:2010aj, Clifton:2011jh, Bajardi:2020fxh}. 

These effective  corrections involve local fields, thus the related theories are described by local actions, according to the principle of locality of the interactions. It is important to distinguish between kinematical and dynamical locality/non-locality. The former refers to the states of the theory: classical theories are kinematically local while quantum theories are kinematically non-local. Instead, the latter refers to the interaction, so that any theory can be dynamically local or non-local depending on the locality/non-locality of the action.

To date, no local ETG is a candidate for a ultimate theory of gravity, mainly because the associated quantum theories are not  fully renormalizable and unitary. 

An  attempt to overcome GR shortcomings is based on the breaking of  locality principle by means of dynamically non-local ETGs. In the last years non-local theories of gravity spread out in the context of Quantum Gravity, especially due to their capability to provide a quantum description for the gravitational interaction. It is a matter of fact that dynamical non-locality is a property shared by all the other fundamental interactions when their one-loop effective actions are considered \cite{Barvinsky:2014lja}. In fact, renormalizable one-loop effective Lagrangians arise after integrating out  massive fermions. This is the case of Quantum Electrodynamics. Here, non-locality is due to the intrinsic non-local nature of  integration, which can be recast as the inverse of a differential operator. Another significant example is provided by the Yukawa theory with a massive scalar field $\phi$, where the non-locality is due to the integral operator $(\square+m^2)^{-1}$. See, \emph{e.g.}  \cite{Yukawa}.

Non-local ETGs  bring  non-locality in  the gravitational interaction, \textit{i.e.} they describe  gravity by non-local effective actions. In general, two  classes of non-local ETGs can  be considered: Infinite Derivative Theories of Gravity (IDGs) and Integral Kernel Theories of Gravity (IKGs). 

IDGs consider analytic transcendental functions of some differential operator, mainly exponential functions of the covariant d'Alembert operator $\square$. An example of such theories is provided by the action \cite{Modesto:2013ioa}:
\begin{equation}
\mathcal{S}=\frac{\kappa}{2} \int \!d^4x\, \sqrt{-g}\,\biggl(R-G_{\mu\nu}\frac{e^{H(-\square_s)}-1}{\square}R^{\mu\nu}\biggr) \,,
\end{equation}
where $\kappa = \frac{c^4}{8\pi G_N}$ and $H(-\square_s)$ is an entire analytic function of $\square_s\equiv\square/M^2_s$, being $M_s$ a mass/lenght scale. It is worthwhile to note that the integral kernel $1/\square\equiv\square^{-1}$ operator is employed only for dimensional reasons since, after a Taylor expansion of the exponential, the denominator $\square$ cancels out. This theory can cure classical Black Hole and Big Bang singularities as shown in  \cite{Modesto:2011kw, Briscese:2012ys}.

IKGs generally employ integral kernels of differential operators, but mainly the inverse operator $\square^{-1}$ (see \cite{Deser:2007jk}). A straightforward non-local extension of GR is 
\begin{equation}
\mathcal{S}=\frac{\kappa}{2} \int\!d^4x \sqrt{-g}\,R\,\Bigl[1+F\Bigl(\square^{-1}R\Bigr)\Bigr] +\mathcal{S}^{(m)}\,.
\end{equation}
The term $\square^{-1}R$ can account for the late-time cosmic expansion without invoking any Dark Energy. From a fundamental physics  point of view, IDGs  emerges in view to obtain fully renormalizable and unitary quantum gravity models \cite{Biswas:2011ar}.  On the other hand, IKGs are inspired by IR quantum corrections coming from  QFT on curved space-time \cite{Barvinsky:2014lja}.

Here we will consider higher-order curvature IKG models described by the general Lagrangian density $F\bigl(R,\square^{-1}R\bigr)$. This effective theory is a generalization of the IKGs considered in literature so far, \textit{e.g.} in Refs. \cite{Nojiri:2007uq,Odi1,Odi2,Bahamonde:2017sdo}. Since this model is a non-local extension of $F(R)$-gravity, it could  account, in principle,  for both UV and IR quantum corrections at once. Specifically,  it could be useful for  achieving both inflationary behavior  (UV) and the today cosmic acceleration (IR). 

In this paper,  we will study  cosmological solutions of $F\bigl(R,\square^{-1}R\bigr)$ gravity using a spatially flat Friedman-Lema\^itre-Robertson-Walker (FLRW) metric as a cosmological background. The main purpose of the analysis is to select physically relevant cosmological models. Nnon-local  gravity models are selected by the so called Noether Symmetry Approach, whose main aspects are outlined \emph{e.g.} in \cite{Capozziello:1996bi, Dialektopoulos:2018qoe}. As shown in literature (see e.g. \cite{Capozziello:2012iea, Urban:2020lfk, Bajardi:2019zzs, baj1, Bajardi:2020xfj, Bajardi:2021tul, Capozziello:2012hm, Capozziello:2007wc, Bajardi:2020mdp}), the approach allows to reduce  dynamics by the existence of symmetries and to find out, eventually, exact solutions. 

In Sec. \ref{NLETGSECT} we overview the foundations of curvature based non-local theories of gravity.  Sec. \ref{NSA} is devoted to the application of the Noether Symmetry Approach to  non-local ETGs depending on  functions of the scalar curvature and the operator $\Box^{-1} R$.  In Sec. \ref{ECS }, we find analytic cosmological solutions coming from the functions selected by the Noether symmetries.  Sec. \ref{concl} is devoted to  the discussion of the results and the  conclusions. 

\section{ Non-Local Curvature Based Theories of Gravity}
\label{NLETGSECT}
Let us introduce now the main properties of non-local  theories of gravity based on curvature invariants\footnote{As discussed in \cite{Bahamonde:2017sdo}, it is possible to formulate non-local theories of gravity based on other geometric invariants as the torsion scalar considering a teleparallel approach.}. Let us start from IKGs. The most general gravitational action in four dimensions, quadratic in the curvature, which can be made ghost-free, must contain infinite covariant derivatives \cite{Biswas:2005qr, Biswas:2011ar, Biswas:2016etb, Biswas:2016egy}. It reads:
\begin{equation}
\mathcal{S}= \frac{\kappa}{2} \int \!d^4x\, \sqrt{-g}\,\Bigl[R+\alpha\Bigl(RF_1(\square_s)R+R_{\mu\nu}F_2(\square_s)R^{\mu\nu}+R_{\mu\nu\rho\sigma}F_3(\square_s)R^{\mu\nu\rho\sigma}\Bigr)\Bigr]   \,,
\end{equation}
where $\alpha\equiv(M_s)^{-2}$ is a dimensional constant,  $M_s$ a mass/lenght scale and $F_i(\square_s)$ transcendental entire analytic functions of the adimensional covariant d'Alembert operator $\square_s\equiv\square/M^2_s$. Being entire functions, they have no pole on the whole complex plane, preventing the occurrence of ghosts. Furthermore, being analytic functions, they can be generally expressed in terms of  Taylor expansion as:
\begin{equation}
F_i(\square_s)=\sum_{n=0}^\infty f_{i,n}(\square_s)^n\,.
\end{equation}

An interesting class of IKGs can be found in Ref. \cite{Modesto:2013ioa}. The corresponding action reads:
\begin{equation}
\mathcal{S}=\frac{\kappa}{2} \int \!d^4x\, \sqrt{-g}\,\biggl(R-G_{\mu\nu}\frac{e^{H(-\square_s)}-1}{\square}R^{\mu\nu}\biggr) \,,
\end{equation}
where $H(-\square_s)$ is an entire analytic function of $\square_s$. The associated field equations at the order $O(R^2)$ are
\begin{equation}
G_{\mu\nu}+O(R^2)=\frac{1}{\kappa}  \, e^{-H(-\square_s)}\,T_{\mu\nu}^{\,(m)}
\label{FEQ}
\end{equation}
and reduce to GR equations at the zeroth order. In a spherically symmetric space-time, the field equations \eqref{FEQ} yield regular black hole solutions without singularities \cite{Modesto:2011kw}. In cosmology, the theory admits bouncing solutions \cite{Briscese:2012ys}. The mechanism of resolution of the singularity is clearly explained in \cite{Buoninfante:2018xiw, Buoninfante:2018mre}. It basically consists of a non-local smearing of the point-like (Dirac delta) source of the Schwarzschild metric induced by the infinite derivatives. This non-local effect implies that the metric is no longer a vacuum solution. In fact, far from the source, gravity is  well described by GR, but once approaching the non-local region $r<2/M_s$ the smearing effects of the source (induced by the non-locality) start being relevant. In particular, all the curvature scalars turn out to be non-singular into this region, so that no singularity occurs even at the gravitational center. 

Another relevant example can be found in \cite{Briscese:2013lna}, where the authors provide the maximal super-renormalizable and unitary UV-completion of the Starobinsky model, whose action is:
\begin{equation}
\mathcal{S}=\frac{\kappa}{2} \int \!d^4x\, \sqrt{-g}\,\biggl[R-G_{\mu\nu}\frac{V_2^{-1}-1}{\square}R^{\mu\nu}+\frac{1}{2}R\,\frac{V_0^{-1}-V_{2}^{-1}}{\square}\,R\biggr] \,,
\label{ActionST}
\end{equation}
where
\begin{equation}
V_2^{-1}\equiv e^{H_2(-\square_s)}p^{(n_2)}(-\square_s) \,,\quad V_0^{-1}-V_{2}^{-1}\equiv\frac{1}{3}\Bigl[e^{H_0(-\square_s)}(1+\square_s)-e^{H_2(-\square_s)}\Bigr] \,,
\end{equation}
with $H_i$ and $p^{(n_2)}$ being entire analytic functions (for details see \cite{Briscese:2013lna}). Any further extensions of the action \eqref{ActionST} can be proved to be non-unitary. 

IDGs are the ultimate consequence of the higher order UV quantum corrections, as discussed in the introduction. The local corrections come from an expansion around $s=0$ of a Schwinger proper time integral \cite{Birrell:book}, which is therefore valid for small times. On the contrary, to get IR corrections, an expansion around $s\to\infty$ is needed. However, the Schwinger proper time integral is meaningful only when the masses of the matter fields are larger than the potential and the space-time curvature. In the massless limit, the proper time integration becomes divergent for late times ($s\to\infty$). This is due to the perturbative nature of the approach, thus a non-perturbative technique to calculate the Schwinger proper time integral is necessary to overcome the issue. This would allow to take into account both UV and non-local IR effects, for every values of potential, curvature and mass. Such technique can be found in \cite{Barvinsky:2014lja} and provides the following non-perturbative action of some QFT in curved space-time:
\begin{equation}
W_0=-\int\!d^4x \sqrt{-g}\,\Bigl[V(x)+V(x)(\square-V)^{-1}V(x)\Bigr]+\frac{1}{6}\,\Sigma \,,
\end{equation}
where $V(x)$ is the potential of the theory and $\Sigma$ is the following surface term:
\begin{equation}
\begin{split}
\Sigma=\int\!d^4x \sqrt{-g}\,\Bigl\{R&-R_{\mu\nu}\,\square^{-1}G^{\mu\nu}+ %
                                           2^{-1} R\,\bigl(\square^{-1}R^{\mu\nu}\bigr) \square^{-1} R_{\mu\nu}+ \\
             &-R^{\mu\nu}\bigl(\square^{-1} R_{\mu\nu}\bigr) \square^{-1}R+ %
                \bigl(\square^{-1} R^{\alpha\beta}\bigr) \bigl(\nabla_{\!\alpha}\,\square^{-1}R\bigr) \nabla_{\!\beta}\,\square^{-1}R +\\
             &-2\,\bigl(\nabla^{\mu}\,\square^{-1} R^{\nu\alpha}\bigr) \bigl(\nabla_{\!\nu}\,\square^{-1} R_{\mu\alpha}\bigr) \square^{-1}R +\\
             &-2\,\bigl(\square^{-1} R^{\mu\nu}\bigr) \bigl(\nabla_{\!\mu}\,\square^{-1}R^{\alpha\beta}\bigr) %
                 \nabla_{\!\nu}\,\square^{-1} R_{\alpha\beta}+O\bigl[R_{\mu \nu}^{\,\,4}\bigr]\Bigr\} \,.
\end{split}
\end{equation}
Further details can be found in \cite{Barvinsky:2014lja}. It is worth stressing that the late-time effective action strongly depends on the integral operator $\square^{-1}$, which is thus able to grasp late-time quantum corrections to GR.
As mentioned above, it was  suggested  in \cite{Deser:2007jk}. The considered action is
\begin{equation}
\label{NL:8}
\mathcal{S}=\frac{\kappa}{2} \int\!d^4x \sqrt{-g}\,R\,\Bigl[1+F\Bigl(\square^{-1}R\Bigr)\Bigr] +\mathcal{S}^{(m)}\,,
\end{equation}
with $F\Bigl(\square^{-1}R\Bigr)$ being an arbitrary function of $\square^{-1}R$. The associated field equations are
\begin{equation}
G_{\mu\nu}+\Delta G_{\mu\nu}= \frac{1}{\kappa}  T_{\mu\nu}^{\,(m)} \,,
\end{equation}
where
\begin{equation}
\Delta G_{\mu\nu}=\,\Bigl(G_{\mu\nu}+g_{\mu\nu}\,\square-\nabla_{\mu}\nabla_{\nu}\Bigr)%
                                \biggl\{F+\square^{-1}\Bigl[R\,F'\Bigr]\biggr\}
                             +\biggl[\delta_{\mu}^{\,\,(\rho}\,\delta_{\nu}^{\,\,\,\sigma)}-\frac{1}{2}\,g_{\mu\nu}g^{\rho\sigma}\biggr]%
                                \partial_{\rho}\Bigl(\square^{-1}R\Bigr)\,\partial_{\sigma}\biggl(\square^{-1}\Bigl[R\,F'\Bigr]\biggr)                  \,,
\end{equation}
with the definitions $F\equiv F\Bigl(\square^{-1}R\Bigr)$ and $\displaystyle{F'\equiv\frac{\partial F}{\partial\bigl(\square^{-1}R\bigr)}}$ .

It can be showed that the operator $\square^{-1}$ can naturally trigger the current late-time cosmic acceleration. Therefore the non-local quantity $\Box^{-1} R$ generates the large numbers required by the current cosmic acceleration avoiding the fine tuning of parameters. Note that these corrections only occur at late-times: during the radiation dominated era the Ricci scalar vanishes and the non-local effects are thus negligible. In general, non-local ETGs are  considered in  literature to address shortcomings of GR in cosmological \cite{Nojiri:2007uq, Jhingan:2008ym, Elizalde:2018qbm, Capozziello:2008gu} and spherically symmetric \cite{Dialektopoulos:2018iph, Capozziello:2021maz} backgrounds.  Specifically, in the latter references, the authors take into account non-local corrections to the Newtonian potential and check how these additional terms can be detected at Galactic scales \cite{Dialektopoulos:2018iph} or at scales of galaxy clusters. Also gravitational waves, coming from non-local gravity, have been considered  \cite{Capriolo1,Capriolo2}.

\section{Noether Symmetry Approach for $F\bigl(R,\square^{-1}R\bigr)$ Gravity}
\label{NSA}
Let us now introduce a class of non-local IKG models which we want to select by the existence of Noether symmetries. The approach can be seen as a physical criterion to select viable models. For a discussion see \cite{Capozziello:2012hm}.

The starting action is:
\begin{equation}
\label{NL:22}
\mathcal{S}=\int\!d^4x \sqrt{-g}\,F\bigl(R,\square^{-1}R\bigr)\,.
\end{equation}
This effective theory is a generalization of $F(R)$-gravity including non-local terms. A point-like Lagrangian, useful for cosmological considerations, can be constructed by the auxiliary \textit{local} scalar field $\phi$, defined as:
\begin{equation}
\label{NL:23}
\phi\equiv\square^{-1}R \quad \textup{or} \quad  R\equiv\square\phi \,.
\end{equation}
In such a way, the action can be "localized" and the starting model can be recast in terms of a  scalar-tensor theory, described by the action
\begin{equation}
\label{NL:24}
S=\int d^4x\,\sqrt{-g}\,\,F(R,\phi).
\end{equation}
Using the Lagrange multipliers method in a FLRW background (with Lagrange Multipliers $\lambda_1$ and $\lambda_2$) both with the cosmological expressions of the Ricci scalar and the higher-order term $\Box \phi$, it is possible to recast the action as
\begin{equation}
\label{NL:27}
S=2\pi^2 \int\!dt\,a^3\biggr\{F(R,\phi)-\lambda_1(R-\ddot{\phi}-3H\dot{\phi})-\lambda_2 \biggl[R+6\biggl(\frac{\ddot{a}}{a}+\Bigl(\frac{\dot{a}}{a}\Bigr)^{\!2}\biggr) \biggr]\biggr\}.
\end{equation}
By varying the action with respect to $R$ we immediately find
\begin{equation}
\lambda_2=\frac{\partial F(R,\phi)}{\partial R}-\lambda_1,
\end{equation}
thus, by promoting $\lambda_1$ to a scalar field and setting $\lambda_1\equiv\lambda(t)$, Eq.~\eqref{NL:27} can be recast as:
\begin{equation}
\label{NL:30}
S=\!\int\!dt\,a^3 \biggr\{F(R,\phi)-\lambda(R-\ddot{\phi}-3H\dot{\phi})-\biggl(\frac{\partial F(R,\phi)}{\partial R}-\lambda\biggr)\biggl[R+6\biggl(\frac{\ddot{a}}{a}+\Bigl(\frac{\dot{a}}{a}\Bigr)^{\!2}\biggr) \biggl]\biggr\}.
\end{equation}
After integrating out the second derivatives, the cosmological point-like  Lagrangian in the  minisuperspace $\mathcal{Q}\equiv\{a, R, \phi, \lambda\}$ reads as 
\begin{equation}
L=\,\,a^3F-a^3\dot{\phi}\dot{\lambda}-a^3R\,\partial_RF+6a\dot{a}^2\partial_RF\,
    -6a\dot{a}^2\lambda+6a^2\dot{a}\dot{R}\,\partial_{RR}F+6a^2\dot{a}\dot{\phi}\,\partial_{R\phi}F-6a^2\dot{a}\dot{\lambda},
    \label{lagrapoint}
\end{equation}
where $F\equiv F(R,\phi)$ and the subscript $R$ denotes the derivative with respect to $R$. Eq. \eqref{lagrapoint} is the  point-like Lagrangian that will be taken into account for the application of the Noether approach discussed in details in the Appendices. 

In the above  minisuperspace $\mathcal{Q}$ of configurations, the first prolongation of the Noether vector reads
\begin{equation}
\label{eqn:X[1]}
X^{[1]}=\,\,\alpha\frac{\partial}{\partial a}+\beta\frac{\partial}{\partial R}+\gamma\frac{\partial}{\partial\phi}+\delta\frac{\partial}{\partial\lambda}
              +(\dot{\alpha}-\dot{\xi}\dot{a})\frac{\partial}{\partial\dot{a}}+%
                                                              (\dot{\beta}-\dot{\xi}\dot{R})\frac{\partial}{\partial\dot{R}}+%
                                                              (\dot{\gamma}-\dot{\xi}\dot{\phi})\frac{\partial}{\partial\dot{\phi}}+%
                                                              (\dot{\delta}-\dot{\xi}\dot{\lambda})\frac{\partial}{\partial\dot{\lambda}} \,.
\end{equation}                                                           
Imposing the existence of Noether symmetry 
\begin{equation}
X^{[1]}L+L\,\dot{\xi}=\dot{g},
\end{equation}
we obtain a system of 28 PDEs, listed in App. \ref{NPSC}. Neglecting linear combinations, the system reduces to six differential equations:

\begin{eqnarray}
&&\begin{split}
     \alpha\,\partial_RF&-\alpha\lambda+a\beta\,\partial_{RR}F+a\gamma\,\partial_{R\phi}F-a\delta+2a\,\partial_RF\,\partial_a\alpha%
     -2a\lambda\,\partial_a\alpha+\\&+a^2\partial_{RR}F\,\partial_a\beta+a^2\partial_{R\phi}F\,\partial_a\gamma%
     -a^2\partial_a\delta-a\,\partial_RF\,\partial_t\xi+a\lambda\,\partial_t\xi=0 \qquad\qquad                   \nonumber              
     \end{split}                                                                                                                                                                             \\
&&\begin{split} 
     2\alpha\,\partial_{RR}F+a\beta\,\partial_{RRR}F+a\gamma\,\partial_{RR\phi}F&+a\,\partial_a\alpha\,\partial_{RR}F%
     +\\&+a\,\partial_R\beta\,\partial_{RR}F-a\,\partial_t\xi\,\partial_{RR}F=0                             \nonumber 
     \end{split}                                                                                                                                                             \\
&&\begin{split}
     12\alpha\,\partial_{R\phi}F&+6a\beta\,\partial_{RR\phi}F+6a\gamma\,\partial_{R\phi\phi}F%
     +6a\,\partial_a\alpha\,\partial_{R\phi}F+\\&+6a\,\partial_{\phi}\beta\,\partial_{RR}F+6a\,\partial_{\phi}\gamma\,\partial_{R\phi}F%
     -a^2\partial_a\delta-6a\,\partial_{R\phi}F\,\partial_t\xi=0                                                         \nonumber                   
     \end{split}                                                                                                                                                                               \\
&&-12\alpha-6a\,\partial_a\alpha+ 6a\,\partial_{\lambda}\beta\,\partial_{RR}F-a^2\partial_a\gamma \nonumber
     -6a\,\partial_{\lambda}\delta+6a\,\partial_t\xi=0                                                                                           \\
&&-3\alpha-a\,\partial_{\phi}\gamma- a\,\partial_{\lambda}\delta+a\,\partial_t\xi=0           \nonumber  \\
&&\begin{split}
     3\alpha F-3\alpha R\,\partial_RF-aR\,\beta\,\partial_{RR}F+a\gamma\,\partial_{\phi}F&-aR\,\gamma\,\partial_{R\phi}F+%
     \\&+ a F\,\partial_t\xi-aR\,\partial_RF\,\partial_t\xi=0 \nonumber.                              
     \end{split}
      \\ \nonumber
     \\ \label{System5} 
\end{eqnarray}

The above system admits the following solution for the infinitesimal generator
\begin{equation}
\xi(t)=(3\tilde{k}_1+\tilde{c}_3)t+k_2,%
\quad \alpha(a)=\tilde{k}_1a,\quad \beta=-2(\tilde{c}_3+3\tilde{k}_1)R,\quad \gamma=c_2 ,\quad \delta(\lambda)=\tilde{c}_3\lambda+\tilde{c}_3 \tilde{c}_1 \,,
\end{equation}
with $\tilde{c}_i, \tilde{k}_i$ constants. Moreover, it turns out that two different functions $F(R,\phi)$ are selected by the Noether symmetries and both of them correspond to the above generator. They read:
\begin{eqnarray}
&&F_I(R,\phi)=-\tilde{c}_1R+[2(\tilde{c}_3+3\tilde{k}_1)R]^{1-\frac{\tilde{c}_3}{2(\tilde{c}_3+3\tilde{k}_1)}}%
                                       \mathcal{F}\biggl(\phi+\frac{c_2\log[2(\tilde{c}_3+3\tilde{k}_1)R]}{2(\tilde{c}_3+3\tilde{k}_1)}\biggr)\,,   \label{eqn:49}
\\
&&F_{II}(R,\phi)=-\tilde{c}_1R+G(R) \, e^{\frac{\tilde{c}_3}{c_2}\phi} \label{secondfunc49}
\end{eqnarray}
where $\displaystyle{\mathcal{F}\biggl(\phi+\frac{c_2\log[2(\tilde{c}_3+3\tilde{k}_1)R]}{2(\tilde{c}_3+3\tilde{k}_1)}\biggr)}$ is an arbitrary integration function which depends on the argument  $\displaystyle{\biggl(\phi+\frac{c_2\log[2(\tilde{c}_3+3\tilde{k}_1)R]}{2(\tilde{c}_3+3\tilde{k}_1)}\biggr)}$.
The first function is a solution of the system \eqref{System5} if and only if the condition $\tilde{c}_3+3\tilde{k}_1\ne0$ holds, while the second relies on the condition $\tilde{c}_3+3\tilde{k}_1= 0$.
These results show, in a straightforward way, how the Noether symmetries select models.

For the sake of simplicity, let us set $m\equiv2(\tilde{c}_3+3\tilde{k}_1)$, so that $F_I$ can be rewritten as
\begin{equation}
\label{modello1}
F_I(R,\phi)=-\tilde{c}_1R+m^{1-\frac{\tilde{c}_3}{m}} R^{1-\frac{\tilde{c}_3}{m}} \mathcal{F}\biggl(\phi+\frac{c_2\log(mR)}{m}\biggr).
\end{equation}
In order to get exact cosmological solutions, the yet unknown function $\mathcal{F}$ must be  carefully chosen. Considering that all those functions which depend on the argument $\displaystyle{\biggl(\phi+\frac{c_2\log[2(\tilde{c}_3+3\tilde{k}_1)R]}{2(\tilde{c}_3+3\tilde{k}_1)}\biggr)}$ admit  Noether symmetries, we consider the simplest choice, namely:
\begin{align*}
&\mathcal{F}_1\biggl(\phi+\frac{c_2\log(mR)}{m}\biggr)\equiv\phi+\frac{c_2\log(mR)}{m}+k \,,\\
\end{align*}
with $k$ being any arbitrary constant. Under this assumption, we get
\begin{align}
&F_1(R,\phi)=-\tilde{c}_1R+k\,m^{1-\frac{\tilde{c}_3}{m}} R^{1-\frac{\tilde{c}_3}{m}}+%
                                           m^{1-\frac{\tilde{c}_3}{m}} R^{1-\frac{\tilde{c}_3}{m}} \phi+%
                                           m^{1-\frac{\tilde{c}_3}{m}} R^{1-\frac{\tilde{c}_3}{m}} \, \frac{c_2\log(mR)}{m} \label{eqn:57-1} \\            
\end{align}
The model $F_1$ is particularly interesting because, by setting $\tilde{c}_3/m=-1$, it reduces to
\begin{equation}
\label{Starobnonlocal}
F_1(R,\phi)\Bigl|_{{\frac{\tilde{c}_3}{m}}=-1}=-\tilde{c}_1R+k\,m^2 R^2+m^2 R^2 \, \phi+c_2 m \, R^2 \log(mR) \,,
\end{equation}
which represents a non-local extension of the Starobinsky cosmological  model. 

Regarding the second function $F_{II}$, by setting $G(R)=k R^n$, with $k, n$ real constants, we obtain the s model
\begin{equation}
\label{eqn:59'}
F_2(R,\phi)=-\tilde{c}_1R+k R^n \, e^{\frac{\tilde{c}_3}{c_2}\phi} \,,
\end{equation}
which is a slight generalization of the model considered in \cite{Bahamonde:2017sdo}. The model $F_2$ indeed, contains the exponential non-local factor appearing in the super-renormalizable and unitary IDGs discussed in Refs. \cite{Modesto:2013ioa, Briscese:2013lna}. In other words, the existence of symmetries selects super-renormalizable models.

Notice that both $F_1(R,\phi)$ and $F_2(R,\phi)$ contain higher order curvature invariants and local scalar fields, which can trigger, in principle,  the early-time inflation and the late-time cosmic acceleration by means of extra geometric terms.

\section{ Cosmological Solutions}
\label{ECS }
We write now the cosmological Euler-Lagrange equations associated to a general $F(R,\phi)$ model. Then we replace into the system the functions selected by the  Noether symmetry  and find out the corresponding exact cosmological solutions. The equations of motion coming from the Lagrangian \eqref{lagrapoint} yield a system of five differential equations:

\begin{subequations}
\renewcommand{\theequation}{\theparentequation.\arabic{equation}}
\begin{eqnarray}
&&\frac{1}{2}F(R,\phi)-\frac{1}{2}\,\dot{\phi}\dot{\lambda}+(\dot{H}+3H^2)\,\partial_RF+(2\dot{H}+3H^2)\lambda%
      +\biggl(2H\frac{d}{dt}+\frac{d^2}{dt^2}\biggr)(-\partial_RF+\lambda)=0   \label{el_a}            \\
&&R=-6\,(2H^2+\dot{H})                                                                                   \label{el_R}           \\
&&\ddot{\phi}+3H\dot{\phi}+12H^2+6\dot{H}=0                                              \label{el_phi}        \\
&&\ddot{\lambda}+3H\dot{\lambda}+\partial_{\phi}F=0                                      \label{el_lambda} \\
&&\frac{1}{2}F(R,\phi)+\frac{1}{2}\,\dot{\phi}\dot{\lambda}+\biggl(3H^2+3H\frac{d}{dt}\biggr)\lambda%
     +\biggl[3(\dot{H}+H^2)-3H\frac{d}{dt}\biggr]\partial_RF=0 \qquad                 \label{energy}
\end{eqnarray}
\end{subequations}
The last equation is the energy condition $E_{ L}\equiv\dot{q}^i \frac{\partial L}{\partial q^i} - L$, which corresponds to the (0,0) component of the field equations. Eqs. \eqref{el_R}, \eqref{el_phi} and \eqref{el_lambda} are the cosmological expression of $R$ and the two Klein-Gordon equations for the scalar fields $\phi$ and $\lambda$, respectively. The Euler-Lagrange equation with respect to the scale factor is the cosmological  Friedmann equation.  

By replacing the first function $F_1(R,\phi)$ into eqs \eqref{el_a}-\eqref{energy},  it is possible to find  the following sets of  solutions:
\\
I):

\begin{eqnarray*}
&& a(t)=a_0\,e^{\Lambda t} \quad R(t)=-12\,\Lambda^2 \quad \phi(t)=-\frac{1}{3} (40+3 k) - 4\Lambda t \quad \lambda(t)=576 m^3 \Lambda^5 t- \frac{C_3 e^{-3\Lambda t}}{3 \Lambda}-\tilde{c}_1,
\\ \text{with} \nonumber
\\
&&  \displaystyle \Lambda=\sqrt{-\frac{1}{12me}} \, (m < 0), \,\,\,\, \tilde{c}_3 = -2m.
\end{eqnarray*}
\\
II)

\begin{eqnarray*}
a(t)=a_0\,t^{\frac{1}{2}} \quad R(t)=0  \quad \phi(t)=C_2\quad \lambda(t)=-\tilde{c}_1-\frac{2 C_3}{\sqrt{t}}.
\end{eqnarray*}
In this case, the equations of motion provide further constraints to the form of the function $F_1(R,\phi)$, which turns out to be reduced with respect to Eq. \eqref{eqn:57-1}. One possible solution is given by GR minimally coupled to a scalar field, namely
\begin{eqnarray*}
F_{1}(R,\phi)=-\tilde{c}_1 R+\phi \,, \\
\end{eqnarray*}
while, in the other case, the free parameters are constrained such that the function takes the form
\begin{equation}
F_1(R,\phi) = - \tilde{c_1} R +  4 \sqrt{2} (- \tilde{c_3})^\frac{5}{4} R^\frac{5}{4} (\phi + k).
\end{equation}
\\
III)
Finally, we have

\begin{eqnarray*}
a(t)=a_0\,t^{-10} \quad R(t) \sim \,t^{-2} \quad \phi(t) \sim C_2 + \log (t) \quad \lambda(t)=-\tilde{c}_1+C_3\,t^{31}+C_4 m^3\,t^{-4}\,.
\end{eqnarray*}
 Specifically, it turns out that the only function associated to solutions I) and III), which contains symmetries and is compatible with the system in Eqs. \eqref{el_a}-\eqref{energy} reads
\begin{equation}
\bar{F}_1(R,\phi)=-\tilde{c}_1R+k m^3 R^3+ m^3 R^3 \phi+ m^3 R^3 \, \frac{c_2\log(mR)}{m} \,.
\end{equation}

Replacing the second function (given by Eq. \eqref{eqn:59'}) into the system \eqref{el_a}-\eqref{energy}, both exponential and power-law solutions occur. As before, depending on the solution considered, the function $F_2$ turns out to be further constrained by the equations of motion, with the result that the integration constants are constrained according to given additional relations (see App. \ref{REL}). The set of solutions reads as:
\\
I)
\begin{eqnarray*}
&&a(t)=a_0\,e^{\Lambda  t}  \\
&&R(t)=-12\Lambda^{2}      \\
&&\phi(t)=-4\Lambda t+C_2\,, \\
&&\lambda(t)=-\frac{3^{n} 4^{n-1} c_{2} k\left(-\Lambda^{2}\right)^{n-1} e^{\frac{C_{2} \tilde{c}_{3}}{c_{2}}-\frac{4 \Lambda \bar{c}_{3}}{c_{2}} t}}{3 c_{2}-4 \tilde{c}_{3}}-\frac{C_{3} e^{-3 \Lambda t}}{3 \Lambda}-\tilde{c}_1, \qquad \qquad \qquad \qquad \qquad \\
&& n =\frac{3 c_2}{4 \tilde{c}_3+c_2}\,, 
\end{eqnarray*}

II)
\begin{eqnarray*}
&&a(t)=a_0\,t^p,  \\
&&R(t)=-6\left(-\frac{p}{t^2}+\frac{2 p^2}{t^2}\right),     \\
&&\phi(t)=C_2-\frac{6 p (2 p-1) \log(t)}{3 p-1},\\
&&\lambda(t)=\frac{6^n (3 p-1)\,ks\, e^{C_2 s} \left(\frac{(1-2 p)p}{t^2}\right)^n t^{2-\frac{6 p (2 p-1) s}{3   p-1}}}%
                   {\left(-2 n-\frac{6 p (2 p-1) s}{3 p-1}+2\right)\left(n (6 p-2)+3 p^2 (4 s-3)-6 p s+1\right)}+%
                    \frac{C_3 \, t^{1-3 p}}{1-3 p} - \tilde{c}_1 \,
                    \\ \\
&& s = \frac{\tilde{c}_3}{\tilde{c}_2}.
\end{eqnarray*}
In the latter case, the parameters $n,p,s$ are dependent according to the relations provided in App. \ref{REL}. 

As an example, by setting $n=1$ from the beginning, the two solutions take the form:
\\
\\
I)$_{n=1}$
\begin{eqnarray*}
&&a(t)=a_0\,e^{\Lambda  t}  \\
&&R(t)=-12\Lambda^{2}      \\
&&\phi(t)=-4\Lambda t+C_2\\
&&\lambda(t)=-3 k\,e^{\frac{C_2}{2}-2 \Lambda  t}-\frac{C_3\,e^{-3 \Lambda t}}{3 \Lambda }- \tilde{c}_1 \qquad \qquad \qquad \qquad \qquad \qquad \qquad \qquad \qquad \qquad 
\end{eqnarray*}
corresponding to the model:
\begin{equation}
\bar{F}_3(R,\phi)=-\tilde{c}_1R+k R\,e^{\frac{\phi}{2}}\,.
\end{equation}
\\
II)$_{n=1}$

\begin{eqnarray*}
&&a(t)=a\, t^p \\
&&R(t)=-6\left(-\frac{p}{t^2}+\frac{2 p^2}{t^2}\right)  \\
&&\phi(t)=C_2-\frac{6 p (2 p-1) \log(t)}{3 p-1} \\
&&\lambda(t)=-\tilde{c}_1-\frac{k (3 p-1) e^{\frac{C_2 (3 p-1)}{6 p-3}}t^{-2 p}}{p-1}+\frac{C_3 t^{1-3 p}}{1-3 p} \qquad \qquad \qquad \qquad \qquad  \qquad \qquad \qquad  \\ \\
&& s=\frac{3 p-1}{3 (2 p-1)} \neq 0, \frac{2}{3},
\end{eqnarray*}
corresponding to the model
\begin{equation}
\bar{F}_3(R,\phi)=-\tilde{c}_1R+k R  \, e^{\frac{3 p-1}{3 (2 p-1)} \phi}\,.
\end{equation}
It is worth noticing that all physically interesting cosmological behaviors can be recovered, in particular accelerating behaviors. They are strictly related to the existence of the symmetry that  allows to reduce dynamics selecting the form of the interacting Lagrangian.

\section{Discussion and Conclusions}
\label{concl}
We considered  effective higher-order IKG models described by the function $F(R,\square^{-1}R)$, a non-local straightforward extension of $F(R)$ gravity,  showing that these models can potentially allow both early and late-time  accelerated   expansion. In particular, inflation would be triggered by higher-order curvature invariants, while late-time expansion by  non-local scalar field $\phi \equiv \square^{-1}R$. This prescription allows to interpret dark energy as a geometric contribution, without introducing any exotic fluid which, to date, has never been observed directly.
 
 We used as a    criterion to select viable models  the existence of Noether symmetries for point-like Lagrangians describing cosmological dynamics. The Euler-Lagrange equations, associated to the symmetries,  yield first integrals of motion which  allow to reduce dynamics and, eventually, to find out  exact solutions. 

A key step towards the search for  symmetries in non-local theories, is the introduction of the auxiliary local scalar field $\phi\equiv\square^{-1}R $. It implements a formal ``localization process'' for the field $\square^{-1}R$\,, so that the theory can be recast in terms of a local  scalar-tensor theory $F(R,\phi)$, with the constraint $\square\phi=R$. 

Noether Symmetry Approach can be applied to the point-like Lagrangian, by using the   existence condition \eqref{nsa1.27}, which leads to a system of 28 PDEs.  We selected two different models containing symmetries and studied the associated cosmological behaviors. 
Both  exponential  and  power-law solutions occur for the scale factor and the applicability to realistic cosmological behaviors depends on the energy ranges related to the parameters.

As a general remark, it is clear that local and non-local contributions work at different scales and this could be a considerable input to address parameter tensions in cosmological behaviour recently reported, in particular the
 $H_0$ tension, namely the discrepancy in the value of the Hubble parameter as obtained from Cosmic Microwave Background data and kinematic measurements related to Cepheids and Supernovae. This issue,  not due to  systematic experimental errors, can represent a real weakness of the $\Lambda$CDM model. A possible explanation for such an incompatibility is to include further degrees of freedom in the gravitational sector. These further degrees of freedom,   coming from local modifications of GR (see \emph{e.g.} \cite{Abadi:2020hbr, Braglia:2020auw, DeFelice:2020cpt}) or non-local modifications \cite{Capozziello:2020nyq}, can alleviate the $H_0$ and the other tensions. In particular, the combination of local and non-local  corrections could address UV and IR behaviors of cosmic history.  

Specifically, the two general models~\eqref{eqn:57-1} and~\eqref{eqn:59'} can be seen as viable effective non-local ETGs matching the different cosmological behaviors.
In a forthcoming study, the above results will be matched with observations.

\section*{Acknowledgments}
F.~B and S.~C. acknowledge the support of {\it Istituto Nazionale di Fisica Nucleare} (INFN) ({\it iniziative specifiche} GINGER, MOONLIGHT2, and QGSKY).

\appendix

\section{Noether Symmetry Approach}

Noether symmetries of the Lagrangian are useful to reduce dynamics and analytically solve systems of differential equations. In what follows we briefly introduce the formulation of the Noether Symmetry Approach used in Sec. \ref{NSA}.

Let us consider the point transformation $(x,y) \to (\bar{x},\bar{y})$ and let $\varepsilon$ be an arbitrary real parameter such that
\begin{equation}
\label{nsa1.2}
\bar{x}=\bar{x}(x,y;\varepsilon),  \qquad \bar{y}=\bar{y}(x,y;\varepsilon).
\end{equation} 
The first-order Taylor expansion of the infinitesimal transformation \eqref{nsa1.2} around $\varepsilon=0$ yields
\begin{align}
&\bar{x}(x,y;\varepsilon)=x+\varepsilon\frac{\partial\bar{x}}{\partial\varepsilon}\bigg|_{\varepsilon=0}  %
                                       =x+\varepsilon\xi(x,y)   \label{nsa1.3} \\
&\bar{y}(x,y;\varepsilon)=y+\varepsilon\frac{\partial\bar{y}}{\partial\varepsilon}\bigg|_{\varepsilon=0}  %
                                       =y+\varepsilon\eta(x,y)  \label{nsa1.4} \,.
\end{align}
The functions $\xi(x,y),\eta(x,y)$ are the components of the tangent vector $\mathbf{X}$ to the orbit of the transformation at the point $(x,y)$, \textit{i.e.}
\begin{equation}
\label{nsa1.5}
\mathbf{X}=\xi(x,y)\frac{\partial}{\partial x}+\eta(x,y)\frac{\partial}{\partial y}.
\end{equation}
Since $(x,y)$ is an arbitrary point, Eq.~\eqref{nsa1.5} provides the tangent vector field to the group orbits, the so called infinitesimal generator of the one-parameter group of point transformations. The prolongation of the tangent vector, involving the $n-th$ derivatives, can be computed by means of the following relations:

\begin{align}
&\bar{y}' \equiv \frac{d\bar{y}(x,y;\varepsilon)}{d\bar{x}(x,y;\varepsilon)}%
                 =\frac{y'(\partial\bar{y}/\partial y)+(\partial\bar{y}/\partial x)}{y'(\partial\bar{x}/\partial y)+(\partial\bar{x}/\partial x)}%
                 =\bar{y}'(x,y,y';\varepsilon),    \label{nsa1.6}  \\              
&\bar{y}''\equiv \frac{d\bar{y}'}{d\bar{x}}=\bar{y}''(x,y,y',y'';\varepsilon), \label{nsa1.7} \\
&\ldots \notag
\end{align}
The prolongations of the generator $\mathbf{X}$ can be obtained through a first-order Taylor expansion around $\varepsilon=0$. Replacing Eqs. \eqref{nsa1.3} and \eqref{nsa1.4} into Eqs. \eqref{nsa1.6} and \eqref{nsa1.7}, the $n^{th}$ derivatives of the transformed coordinates (up to the first order) read:
\begin{align}
&\bar{y}'=y'+\varepsilon\left(\frac{d\eta}{dx}-y'\frac{d\xi}{dx}\right) =y'+\varepsilon\eta^{[1]}  \,,   \\
&\vdots   \notag\\
&\bar{y}^{(n)}=y^{(n)}+\varepsilon\left(\frac{d\eta^{(n-1)}}{dx}-y^{(n)}\frac{d\xi}{dx}\right)  %
                         =y^{(n)}+\varepsilon\eta^{[n]}  \,,
\end{align}
where
\begin{equation}
\label{nsa1.10}
\eta^{[n]} \equiv \frac{d\eta^{(n-1)}}{dx}-y^{(n)}\frac{d\xi}{dx}=\frac{d^n}{dx^n}(\eta-y'\xi)+y^{(n+1)}\xi \,
\end{equation}
is the $n^{th}$ prolongation function of $\eta$. Therefore, the $n^{th}$ prolongation of Noether's vector can be written as:
\begin{equation}
\label{nsa1.12}
\mathbf{X}^{[n]}=\mathbf{X}+\eta^{[1]}\partial_{y'} +...+\eta^{[n]}\partial_{y^{(n)}}
\end{equation}
Starting from Eq.~\eqref{nsa1.12} and considering a point-like Lagrangian $L=L\left(t,q(t),\dot{q}(t)\right)$, with $q^i$ being the coordinates and $t$ the time, the first prolongation of Noether's vector can be written as
\begin{equation}
\label{nsa1.26}
\mathbf{X}^{[1]}=\xi(t,q)\frac{\partial}{\partial t}+\eta^i(t,q)\frac{\partial}{\partial q^i}%
                                +(\dot{\eta}^i-\dot{\xi}\dot{q}^i) \frac{\partial}{\partial\dot{q}^{i}} \,.
\end{equation}
The first Noether Theorem states that the one-parameter group of point transformations generated by $\mathbf{X}$ is a one-parameter group of Noether point symmetries for the dynamical system described by $L$, if and only if there exists a function $g\bigl(t,q(t)\bigr)$ such that 
\begin{equation}
\label{nsa1.27}
\mathbf{X}^{[1]}L+\dot{\xi}L=\dot{g},
\end{equation}
 whose associated first integral of motion is:
\begin{equation}
\label{nsa1.28}
I(t,q,\dot{q})=\xi \left(\dot{q}^i \frac{\partial L}{\partial\dot{q}^i}-L \right)-\eta^i \frac{\partial L}{\partial\dot{q}^i}+g\,.
\end{equation}
\section{System of Differential Equations Coming from Noether's Symmetry Existence Condition}
\label{NPSC}
The system of differential equations coming from the symmetry existence condition is:
\begin{subequations}
\renewcommand{\theequation}{\theparentequation.\arabic{equation}}
\begin{eqnarray}
&&12a\,\partial_RF\,\partial_t\alpha-12a\lambda\,\partial_t\alpha+6a^2\partial_{RR}F\,\partial_t\beta%
    +6a^2\partial_{R\phi}F\,\partial_t\gamma-6a^2\,\partial_t\delta\,+a^3F\,\partial_a\xi-a^3R\,\partial_RF\,\partial_a\xi=\partial_ag%
    \qquad\qquad\quad                                                                                                                              \label{eqn:19.1}                       \\      
&&6a^2\partial_{RR}F\,\partial_t\alpha+a^3F\,\partial_R\xi-a^3R\,\partial_RF\,\partial_R\xi=\partial_Rg        \label{eqn:19.2}                        \\ &&6a^2\partial_{R\phi}F\,\partial_t\alpha-a^3\partial_t\delta%
    +a^3F\,\partial_{\phi}\xi-a^3R\,\partial_RF\,\partial_{\phi}\xi=\partial_{\phi}g                                               \label{eqn:19.3}                        \\
&&-6a^2\partial_t\alpha-a^3\partial_t\gamma+a^3F\,\partial_{\lambda}\xi%
    -a^3R\,\partial_RF\,\partial_{\lambda}\xi=\partial_{\lambda}g                                                                           \label{eqn:19.4}                       \\
&&\begin{split}
    \alpha\,\partial_RF&-\alpha\lambda+a\beta\,\partial_{RR}F+a\gamma\,\partial_{R\phi}F-a\delta+2a\,\partial_RF\,\partial_a\alpha%
     -2a\lambda\,\partial_a\alpha+ \\&+a^2\partial_{RR}F\,\partial_a\beta+a^2\partial_{R\phi}F\,\partial_a\gamma%
     -a^2\partial_a\delta-a\,\partial_RF\,\partial_t\xi+a\lambda\,\partial_t\xi=0                                                       \label{eqn:19.5}
   \end{split}                                                                                                                                                                                                            \\
&&6a^2\partial_{RR}F\,\partial_R\alpha=0                                                                                                                \label{eqn:19.6}                 \\
&&6a^2\partial_{R\phi}F\,\partial_{\phi}\alpha-a^3\partial_{\phi}\delta=0                                                             \label{eqn:19.7}                 \\
&&6a^2\partial_{\lambda}\alpha+a^3\partial_{\lambda}\gamma=0                                                                          \label{eqn:19.8}               \\
&&6a\,\partial_RF\,\partial_a\xi-6a\lambda\,\partial_a\xi=0                                                                                        \label{eqn:19.9}               \\
&&\begin{split} 
    12a\,\partial_{RR}F\,\alpha&+6a^2\partial_{RRR}F\,\beta+6a^2\partial_{RR\phi}F\,\gamma+6a^2\,\partial_{RR}F\,\partial_a\alpha%
    +12a\,\partial_RF\,\partial_R\alpha+\\&-12a\lambda\,\partial_R\alpha+6a^2\partial_{RR}F\,\partial_R\beta+6a^2\partial_{R\phi}F\,%
    \partial_R\gamma-6a^2\partial_R\delta-6a^2\partial_{RR}F\,\partial_t\xi=0                                                           \label{eqn:19.10}
  \end{split}                                                                                                                                                                                                                \\
&&\begin{split}
    12a\,\partial_{R\phi}F\,\alpha&+6a^2\partial_{RR\phi}F\,\beta+6a^2\partial_{R\phi\phi}F\,\gamma%
    +6a^2\,\partial_{R\phi}F\,\partial_a\alpha+12a\,\partial_RF\,\partial_{\phi}\alpha+\\&-12a\lambda\,\partial_{\phi}\alpha%
    +6a^2\partial_{RR}F\,\partial_{\phi}\beta+6a^2\partial_{R\phi}F\,\partial_{\phi}\gamma%
     -a^3\partial_a\delta-6a^2\partial_{\phi}\delta-6a^2\partial_{R\phi}F\,\partial_t\xi=0                                   \label{eqn:19.11}
  \end{split}                                                                                                                                                                                                               \\
&&\begin{split}
    -12a\alpha-6a^2\partial_a\alpha&+\!12a\,\partial_RF\,\partial_{\lambda}\alpha-\!12a\lambda\,\partial_{\lambda}\alpha%
    \!+\! 6a^2\partial_{RR}F\,\partial_{\lambda}\beta+\\&-a^3\partial_a\gamma+6a^2\,\partial_{R\phi}F\,\partial_{\lambda}\gamma%
     -6a^2\,\partial_{\lambda}\delta+6a^2\,\partial_t\xi=0                                                                                    \label{eqn:19.12}
   \end{split}                                                                                                                                                                                                                \\
&&6a^2\partial_{R\phi}F\,\partial_R\alpha+6a^2\partial_{RR}F\,\partial_{\phi}\alpha-a^3\partial_R\delta=0   \label{eqn:19.13}                      \\
&&-6a^2\partial_R\alpha+6a^2\partial_{RR}F\,\partial_{\lambda}\alpha-a^3\partial_R\gamma=0                     \label{eqn:19.14}                      \\
&&-3a^2\alpha-6a^2\partial_{\phi}\alpha+6a^2\partial_{R\phi}F\,\partial_{\lambda}\alpha-a^3\partial_{\phi}\gamma-%
      a^3\partial_{\lambda}\delta+a^3\partial_t\xi=0                                                                                               \label{eqn:19.15}                      \\
&&6a^3\partial_a\xi-6a^2\partial_{R\phi}F\,\partial_{\lambda}\xi+6a^2\partial_{\phi}\xi=0                                \label{eqn:19.16}                     \\
&&-6a^2\partial_{RR}F\,\partial_{\lambda}\xi+6a^2\partial_R\xi=0                                                                       \label{eqn:19.17}                     \\
&&-6a^2\partial_{RR}F\,\partial_{\phi}\xi-6a^2\partial_{R\phi}F\,\partial_R\xi=0                                                  \label{eqn:19.18}                     \\
&&a^3\partial_R\xi=0                                                                                                                                                 \label{eqn:19.19}                   \\
&&a^3\partial_{\phi}\xi=0                                                                                                                                            \label{eqn:19.20}                   \\
&&a^3\partial_{\lambda}\xi=0                                                                                                                                    \label{eqn:19.21}                    \\
&&6a^2\partial_{\lambda}\xi=0                                                                                                                                   \label{eqn:19.22}                  \\
&&-6a^2\partial_{R\phi}F\,\partial_{\phi}\xi=0                                                                                                           \label{eqn:19.23}                  \\
&&-6a^2\partial_{RR}F\,\partial_R\xi=0                                                                                                                     \label{eqn:19.24}                   \\
&&-6a\,\partial_RF\,\partial_R\xi+6a\lambda\,\partial_R\xi-6a^2\partial_{RR}F\,\partial_a\xi=0                            \label{eqn:19.25}                   \\
&&-6a\,\partial_RF\,\partial_{\phi}\xi+6a\lambda\,\partial_{\phi}\xi-6a^2\partial_{R\phi}F\,\partial_a\xi=0            \label{eqn:19.26}                  \\
&&-6a\,\partial_RF\,\partial_{\lambda}\xi+6a\lambda\,\partial_{\lambda}\xi+6a^2\partial_a\xi=0                           \label{eqn:19.27}                 \\
&&3\alpha F-3R\,\alpha\,\partial_RF-aR\,\beta\,\partial_{RR}F+a\gamma\,\partial_{\phi}F-aR\,\gamma\,\partial_{R\phi}F+%
     aF\,\partial_t\xi-aR\,\partial_RF\,\partial_t\xi=a^{-2}\partial_tg                                                                       \label{eqn:19.28}
\end{eqnarray}
\end{subequations}
\newpage
\section{Relations Among Free Parameters in the Power-Low Solution of the Model $F_2(R,\Phi)$}
\label{REL}
The  two equations which determine the relation among the three free parameters $n,s,p$, corresponding to the model $F_2$ of section \ref{NSA}, are:
\begin{equation}
\label{ecs1.37}
 \begin{split}
 & \Bigl[4 n^4 (3 p-1)^3+2 n^3 (1-3 p)^2 \bigl(6 p^2 (6 s-1)-p (18 s+13)+5\bigr)+\\ 
 &\,\,+n^2 (3 p-1) \Bigl(9 p^4  (48 s^2-16 s-3)-3 p^3  (144 s^2+80 s-33)+\\
 &\,\,+3 p^2  (36 s^2+92 s+5)-3 p (20 s+13)+8\Bigr)+n\Bigl(54 p^6 s (16 s^2-8 s-3)+\\
 &\,\,-9 p^5  (144 s^3+56 s^2-39 s-27)+9 p^4  (72 s^3+132 s^2+31 s-51)+\\
 &\,\,-9 p^3  (12s^3+66 s^2+61 s-27)+3 p^2  (30 s^2+77 s-9t)-10 p (3 s+1)+2\Bigr)+\\
 &\,\,-3 (1-3 p)^2 p (2 p-1)\bigl(3 p^2 s (4 s-3)-p  (6 s^2+s-3)+2 s-1\bigr) \Bigr] =0, \nonumber
\end{split}
\end{equation}

\begin{equation}
\label{ecs1.38}
 \begin{split}
  & \Bigl[2 n^3 (1-3 p)^2+n^2 (3 p-1) \bigl(3 p^2 (8 s+1)-4 p (3 s+4)+5\bigr)+\\
  &\,\,+n \Bigl(18 p^4 s (4 s+1)-3 p^3 (24 s^2+35 s+9)+6 p^2 (3 s^2+13 s+9)-3 p (5 s+9)+4\Bigr)+\\
  &\,\,-(1-3 p)^2 (2 p-1) (3 p s-1)\Bigr] =0.
 \end{split}
 \end{equation}

\end{document}